\documentclass[showpacs,twocolumn,showkeys,amsmath,amssymb,pra,superscriptaddress,nofootinbib]{revtex4-1} 
\usepackage{amsmath}
\usepackage{color}
\usepackage{graphicx}   
\usepackage{dcolumn}    
\usepackage{bm}         

\begin{document} 

\title[]{Bright soliton quantum superpositions: signatures of high- and low-fidelity states}
\author{Bettina Gertjerenken}
\email{b.gertjerenken@uni-oldenburg.de}
\affiliation{Institut f\"ur Physik, Carl von Ossietzky Universit\"at, D-26111 Oldenburg, Germany}

\keywords{bright soliton, mesoscopic entanglement, Bose-Einstein condensation}
                  
\date{\today}
 
\begin{abstract}
Scattering quantum bright solitons off barriers has been predicted to lead to nonlocal quantum superpositions, in particular the NOON-state. The focus of this paper lies on signatures of both high- and low-fidelity quantum superposition states. We numerically demonstrate that a one-dimensional geometry with the barrier potential situated in the middle of an additional -- experimentally typical -- harmonic confinement gives rise to particularly well-observable signatures. In the elastic scattering regime we investigate signatures of NOON-states on the $N$-particle level within an effective potential approach. We show that removing the barrier potential and subsequently recombining both parts of the quantum superposition leads to a high-contrast interference pattern in the center-of-mass coordinate for narrow and broad potential barriers. We demonstrate that the presented signatures can be used to clearly distinguish quantum superpositions states from statistical mixtures and are sufficiently robust against experimentally relevant excitations of the center-of-mass wave function to higher lying oscillator states. For two-particle solitons we extend these considerations to low-fidelity superposition states: even for strong deviations from the two-particle NOON-state we find interference patterns with high contrast.
\end{abstract}

\pacs{03.75.Gg, 03.75.Lm, 34.50.Cx, 67.85.-d}

\maketitle


\section{Introduction}
The experimental realization of mesoscopic entanglement is in the focus of current research: interest stems both from fundamental aspects as tests of decoherence mechanisms~\cite{Zurek03} and the relevance for quantum-enhanced interferometry~\cite{Giovanetti04}. Recent suggestions for realizations of nonlocal mesoscopic superpositions include Bose-Einstein condensates~(BECs)~\cite{CarrEtAl10}, cavity quantum optomechanical systems~\cite{RomeroIsartEtAl11} and topological defects~\cite{DziarmagaEtAl12}. Bright solitons\footnote{In the following the term ``soliton'' is used synonymously with ``solitary wave''.}, self-bound matter-waves generated from Bose-Einstein condensates~\cite{PethickSmith02,Stringari03} are, in their quantum version~\cite{LaiHaus89,CastinHerzog01,SykesEtAl07}, a particularly promising system to generate quantum superpositions~\cite{WeissCastin09,StreltsovEtAl09,GertjerenkenEtAl13}.

Landmark experiments have already realized bright solitons which behave close to Gross-Pitaevskii solitons~\cite{PethickSmith02,HelmEtAl12,CuevasEtAl2013}: both single bright matter-wave solitons~\cite{KhaykovichEtAl02} and soliton trains~\cite{StreckerEtAl02,CornishEtAl06} have been created in \mbox{(quasi-)}one-dimensional attractive Bose gases. Scattering bright solitons off barrier potentials is currently investigated in a new generation of experiments~\cite{PollackEtAl10,Marchant2013}.

The stability of solitons~\cite{KhaykovichEtAl02,BillamMarchant13} makes them a candidate system for matter-wave interferometry~\cite{MartinEtAl12,HelmEtAl12,CuevasEtAl2013}. Theoretical investigations of one-dimensional attractive Bose gases also comprise soliton localization via disorder~\cite{Mueller2011B}, polaritonic solitons in an optical lattice~\cite{MalomedEtAl13}, effects of higher dimension~\cite{BillamMarchant13,CuevasEtAl2013}, macroscopic quantum tunneling~\cite{GlickCarr11}, resonant trapping in a quantum well~\cite{ErnstEtAl10}, collision-induced entanglement of indistinguishable solitons~\cite{LewensteinMalomed09}, the creation of Bell states via collisions of distinguishable solitons~\cite{GertjerenkenEtAl13}, and collision dynamics and entanglement generation of two initially independent and indistinguishable boson pairs~\cite{Holdaway13}.

While a mean-field description has been found to successfully describe many aspects in the scattering dynamics of bright solitons for high kinetic energies~\cite{AkkermansEtAl08,CuevasEtAl2013,Marchant2013} considerably deviating behavior has been predicted for lower kinetic energies. In this regime scattering bright solitons off a barrier potential gives rise to a continuously varying reflection/transmission coefficient on the $N$-particle quantum level while a discontinuous behavior has been observed on the Gross-Pitaevskii level~\cite{WangEtAl12,GertjerenkenBillamEtAl12}. This indicates the formation of quantum superposition states at the $N$-particle level which are not allowed by the nonlinear Gross-Pitaevskii equation~\cite{GertjerenkenBillamEtAl12}.

In~\cite{WeissCastin09,StreltsovEtAl09} elastically scattering bright solitons off a barrier potential has been predicted to lead to nonlocal mesoscopic superposition states
\begin{equation}\label{eq:NOON}
|\Psi_{\mathrm{NOON}}\rangle = \frac{1}{\sqrt{2}}\left(|N,0 \rangle + \mathrm{e}^{\mathrm{i}\phi}|0,N \rangle\right),
\end{equation}
where all the particles are placed in a coherent superposition with a distance between the two parts of the wave function that is much larger than the soliton size. Here, and in the following, the notation~$|n,N-n\rangle$ signifies that~$n$~($N-n$) particles are situated to the left (right) of the barrier potential. The NOON-state is very sensitive to decoherence: a single atom loss suffices to destroy the quantum superposition. In this view particle numbers on the order of $N=100$ have been suggested~\cite{WeissCastin09,StreltsovEtAl09}. Experimental requirements for the realization of the suggested protocols are low temperatures~\cite{LeanhardtEtAl03}, a good vacuum~\cite{AndersonEtAl95} and the particle-number control available in experiments like~\cite{GrossEtAl10}, cf. Sec.~\ref{sec:harmonicconfinement}.

Another interesting class of states, also relevant for quantum-enhanced interferometry, are general superposition states involving contributions like
\begin{equation}\label{eq:lowfidcat}
|\Psi\rangle_n = \frac{1}{\sqrt{2}}\left(|N-n,n\rangle + \mathrm{e}^{\mathrm{i}\phi_n}|n,N-n\rangle \right).
\end{equation}
Such a quantum superposition is less sensitive to decoherence via atom losses: it would be turned into a statistical mixture of still entangled states. Additionally, in the presented protocols low-fidelity quantum superpositions, enabling higher initial CoM kinetic energies, could be realized on shorter time-scales than the NOON-state.

An unambiguous experimental demonstration requires clear, experimentally measurable signatures that distinguish quantum superpositions from statistical mixtures, which is the focus of this work. In~\cite{WeissCastin09} it has been suggested to switch off the scattering potential and let both parts of the quantum superposition recombine and interfere. These interference patterns are particularly well-observable in the center-of-mass (CoM) density~\cite{GertjerenkenEtAl11}
\begin{equation}\label{eq:CoMdensity}
\rho_{\rm{CoM}} (x)= \langle \delta\left(x-X\right) \rangle,
\end{equation}
while in general\footnote{\label{CoMandsingleparticledensity}For particle numbers as low as $N=2$ interference patterns in the single-particle density can be observed but with reduced contrast in comparison to the CoM density. For larger particle numbers only the CoM density gives rise to high-contrast interference patterns, cf.~\cite{GertjerenkenEtAl11}.} they vanish in the single-particle density~\cite{GertjerenkenEtAl11}
\begin{equation}\label{eq:onedensity}
\rho_{\rm{one}}(x) = \frac{1}{N}\sum_{j}\langle \delta\left(x-x_j\right) \rangle.
\end{equation}
Here, $\langle \cdot \rangle$ denotes the expectation value, $x$ the spatial coordinate, $x_j$ the positions of the particles and $X=\frac{1}{N}\sum_{j = 1}^N x_{j}$ the CoM of the system. The CoM has to be treated quantum mechanically~\cite{WeissCastin09,Mueller2011B} and is measurable with larger precision than the width of the soliton. In contrast to the interference of two Bose-condensates where a single experiment yields an interference pattern~\cite{AndrewsEtAl97}, the CoM density~(\ref{eq:CoMdensity}) is determined in a series of measurements: each run gives a single point and an interference patterns builds up as for single-photon interference~\cite{GravierEtAl86}. In the following we assume sufficient experimental stability to guarantee shifts of the interference pattern from run to run to be smaller than the distance of neighboring interference maxima to avoid washing out of the pattern.

Here, we investigate the scattering of a bright soliton on a barrier potential that is located in the middle of an experimentally typical~\cite{StreckerEtAl02,CornishEtAl06,PollackEtAl10,Marchant2013} slight harmonic confinement and show that this set-up leads to advantageous signatures of quantum superposition states. First, we investigate the elastic scattering regime within the effective potential approach~\cite{WeissCastin09,SachaEtAl09,WeissCastin12}. In contrast to previous work~\cite{GertjerenkenBillamEtAl12}, where scattering twice off the barrier potential has been investigated, we focus on interference patterns in the CoM density after scattering once off the barrier potential. While the first has been identified as a clear signature of quantum superpositions particularly suited for not-too-broad effective potentials, we demonstrate high-contrast interference patterns in the CoM density both for narrow and broad effective potentials. We also show that both signatures are sufficiently robust to experimentally relevant excitations in the CoM coordinate for realistic parameters.

An interesting question is if interference patterns with suitable contrast are still observable for general quantum superposition states~(\ref{eq:lowfidcat}), requiring $N$-particle methods beyond the effective potential approach. At the example of a two-particle soliton we demonstrate sufficiently high values of the contrast via discretization of the two-particle Schr\"odinger equation.

The paper is structured as follows: Section~\ref{eq:models} introduces the underlying $N$-particle methods and the numerical implementation via discretization of the Schr\"odinger equation. In Sec.~\ref{sec:harmonicconfinement} experimental requirements for the creation of quantum superposition states in this set-up are described. In Sec.~\ref{eq:highfidelity} the resulting interference patterns in the CoM density are investigated for high-fidelity NOON-states~(\ref{eq:NOON}) within the effective potential approach. A model for the effect of excitations in the CoM coordinate is outlined in Sec.~\ref{sec:excitations}. In Sec.~\ref{eq:generalsup} the discussion is extended to general, low-fidelity quantum superposition states for two-particle solitons. Section~\ref{eq:conclusion} concludes the paper.

\section{Models}\label{eq:models}
The system can be modeled on the $N$-particle level with the exactly solvable~\cite{LaiHaus89,CastinHerzog01,SykesEtAl07} Lieb-Liniger(-McGuire) Hamiltonian~\cite{LiebLiniger63,McGuire64} with additional external potential~$V_{\mathrm{ext}}$:
\begin{align}\label{eq:LiebLiniger}
H = &-\sum_{j=1}^{N}\frac{\hbar^2}{2m}\partial^2_{x_j} + \sum_{j=1}^{N-1}\sum_{n=j+1}^N g_{1\mathrm{D}}\delta\left(x_j-x_n\right)\nonumber \\ &+\sum_{j=1}^{N}V_{\mathrm{ext}}\left(x_j\right),
\end{align}
Here, $N$ denotes the particle number, $m$ the particle mass and attractive contact interaction with coupling constant $g_{1\rm{D}}<0$ is assumed.

For zero external potential eigensolutions of the resulting Schr\"odinger equation are (up to a phase factor) translationally invariant:
\begin{equation}\label{eq:manyparticlesoliton}
\Psi_{N,k}\left(\mathbf{x},t\right) \propto \exp \left(-\beta \sum_{1\leq j < n \leq N} |x_{j} - x_{n}| + \mathrm{i}k \sum_{j=1}^N x_j\right),
\end{equation}
with $\beta = -mg_{1\rm{D}}/2\hbar^2 > 0$. Eigenenergies $E=E_0(N)+E_{\rm{kin}}$ are given by the CoM kinetic energy
\begin{equation}
E_{\rm{kin}} = N\frac{\hbar^2k^2}{2m},
\end{equation}
and the ground state energy~\cite{McGuire64,CastinHerzog01}
\begin{equation}
E_0(N)=-\frac{1}{24}\frac{mg^2_{1\rm{D}}}{\hbar^2}N(N^2-1)
\end{equation}
of the quantum soliton
\begin{equation}\label{eq:groundstate}
\psi_{0}(\mathbf{x}) = C_N\exp\left(-\beta \sum_{1\leq j < n \leq N} |x_{j} - x_{n}| \right),
\end{equation}
with $C_N =\left[\frac{(N-1)!}{N}(2\beta)^{N-1}\right]^{1/2}$~\cite{CastinHerzog01}. It is separated from a continuum of solitonic fragments by the energy gap~\cite{CastinHerzog01}
\begin{align}\label{eq:energygap}
|\mu| & \equiv  E_0(N-1) - E_0(N) \nonumber \\
&= \frac{mg^2_{1\rm{D}}N(N-1)}{8\hbar^2}.
\end{align}
Taking a delta-function~$\delta(X-x_0)$ for the CoM wave function leads in the limit $N\gg1$ to a single-particle density~(\ref{eq:onedensity}) of the quantum soliton~(\ref{eq:groundstate}) identical to the mean-field density~\cite{CalogeroDegasperis75,CastinHerzog01}
\begin{equation}\label{eq:mfsoltiondensity}
\rho_{\rm{mf}}(x) \simeq \frac{N}{2\ell\cosh^2\left((x-x_0)/\ell\right)}
\end{equation}
with
\begin{equation}\label{eq:ell}
\ell = 2\frac{\hbar^2}{m|g_{1\mathrm{D}}|(N-1)}.
\end{equation}
While in this case mean-field and $N$-particle solutions agree, mesoscopic quantum superpositions which are in the focus of this work cannot be described by the nonlinear Gross-Pitaevskii equation.

\subsection{Effective potential approach}\label{sec:effpot}
In the low-energy regime scattering bright solitons off barrier potentials can be described on the $N$-particle level within the mathematically rigorous~\cite{WeissCastin12} effective potential approach~\cite{WeissCastin09,SachaEtAl09}. This approach is particularly suited to obtain physical insight into the scattering dynamics and gives an effective Schr\"odinger equation for the CoM motion (as in~\cite{GertjerenkenBillamEtAl12} with additional harmonic confinement):
\begin{align}\label{eq:CoMSGL}
\mathrm{i}\hbar \partial_t \Psi\left(X,t \right) = & \left[-\frac{\hbar^2}{2Nm}\partial^2_X + \frac{1}{2}Nm\omega^2X^2\right]\Psi\left(X,t\right)\nonumber \\ & +V_{\mathrm{eff}}\left(X\right)\Psi\left(X,t\right).
\end{align}
Here $\omega$ denotes the axial trapping frequency. The effective potential~$V_{\mathrm{eff}}$ is the convolution of the internal density profile of the soliton with the barrier potential~$V(\mathbf{x})$:
\begin{equation}\label{eq:effpot}
V_{\mathrm{eff}}(X) = \int d^Nx|\Psi_{N,k}(\mathbf{x})|^2V(\mathbf{x})\delta\left(X-\frac{1}{N}\sum_{\nu=1}^N x_{\nu}. \right)
\end{equation}
Assuming the barrier potential to be a delta function~$v_0\delta(x)$ and using the results of~\cite{CalogeroDegasperis75} the evaluation of Eq.~(\ref{eq:effpot}) yields
\begin{equation}\label{eq:effpotU0}
V_{\mathrm{eff,c}}(X) = \frac{U_0}{\cosh^2\left(X/\ell\right)} 
\end{equation}
with $\ell$ introduced in equation~(\ref{eq:ell}) and
\begin{equation}
U_0 \equiv \frac{Nv_0}{4}\frac{m|g_{1\mathrm{D}}|(N-1)}{\hbar^2}.
\end{equation} 
The effective potential thus has the form of the soliton. Narrow effective potentials imply small soliton and barrier widths. If the effective potential is narrow enough, it can be approximated with a delta function
\begin{equation}\label{eq:Veffdelta}
V_{\mathrm{eff},\delta} = \frac{\hbar^2}{m}\Omega\delta\left(X + X_{\mathrm{s}}\right).
\end{equation}
where we also allow for shifts~$X_{\mathrm{s}}$ of the barrier potential out of the middle of the harmonic confinement. Broad potentials of the form~(\ref{eq:effpotU0}) would also be obtained by broader scattering potentials as in current set-ups~\cite{Marchant2013}. For the effective potential approach to be valid these have to be sufficiently smooth which for using a laser focus as a scattering potential will always be the case.

\subsection{Numerical implementation}\label{sec:numimp}
The situation can be modeled with a Bose-Hubbard Hamiltonian with additional harmonic confinement,
\begin{eqnarray}\label{eq:BHH}
H_{\rm{discr}} &=&-J\sum_j\left(a_j^{\dagger}a_{j+1}^{\phantom{\dagger}} + a_{j+1}^{\dagger}a_{j} \right) +\frac{U}{2}\sum_{j}n_{j}\left(n_{j} -1 \right)\nonumber \\
&&+A\sum_{j}n_jj^2 + \tilde{v}_0\delta_{j,0}.
\end{eqnarray}
both on the $N$-particle level within the effective potential approach from Sec.~\ref{sec:effpot} (in this case without interaction term) and for a two-particle soliton. Here~$J$ denotes the tunneling strength, $U$ the on-site interaction strength,~$A$ the strength of the harmonic confinement and~$\tilde{v}_0$ the strength of the delta-like barrier potential. The operators~$a_j^{\left(\dagger\right)}$ annihilate (create) a particle at lattice site~$j$, and $n_{j}$ is the particle number operator for lattice site~$j$. The time-evolution corresponding to the Hamiltonian~(\ref{eq:BHH}) is computed via the Shampine-Gordon routine~\cite{ShampineGordon75} for sufficiently small lattice spacing~$b$: in the limit~$b\rightarrow 0$ the Lieb-Liniger model with additional harmonic confinement is recovered.

\section{Scattering bright solitons off barrier potentials in additional harmonic confinement}\label{sec:harmonicconfinement}
In current experiments scattering bright solitons off barrier potentials is investigated in additional harmonic confinement~\cite{PollackEtAl10,Marchant2013}. Here, we model such a protocol for the low-energy regime necessary for the production of quantum superposition states~\cite{WeissCastin09,StreltsovEtAl09}.

Initially the many-particle ground state is prepared in the harmonic trap~\cite{Holdaway12}: throughout this work, we assume that the internal degrees of freedom are described by the Lieb-Liniger soliton~(\ref{eq:groundstate}) while the CoM motion is determined by the harmonic confinement. The center of the trap is then (quasi-)instantaneously shifted and the scattering potential in the middle of the trap is switched on.

In the following we specify the experimental requirements necessary for the creation of mesoscopic quantum superpositions. The initial state has to be prepared carefully: With probability
\begin{equation}\label{eq:stateocc}
p_{\mathrm{CoM,gr}} \simeq 1- \exp\left(-\frac{\hbar \omega}{k_{\mathrm{B}}T}\right).
\end{equation}
the initial CoM wave function is given by the ground state of the harmonic oscillator. To avoid excitations to higher lying oscillator states temperatures in the range of 450~pK, as in~\cite{LeanhardtEtAl03}, are required. The temperature should also be small in comparison with~$|\mu|$, as defined in Eq.~(\ref{eq:energygap}), to avoid excitations of single particles out of the soliton. Additionally, to ensure elastic scattering we assume low CoM kinetic energies~\cite{WeissCastin09,GertjerenkenBillamEtAl12}, corresponding to weak harmonic confinement. To reduce decoherence by single-particle losses a very good vacuum~\cite{AndersonEtAl95} is required. No thermal rest gas should be present in the harmonic trap. To ensure clear signatures a particle number post selection as in~\cite{GrossEtAl10} is assumed.

Experimentally realistic parameters~\cite{WeissCastin09} are\footnote{In~\cite{WeissCastin09} a slightly larger axial trapping frequency~$\omega = 2\pi\cdot 23.5$~Hz was chosen. In contrast to this work, the axial trapping potential is used for the initial state preparation and subsequently opened.}
\begin{equation}\label{eq:expparam}
\omega = 2\pi\cdot10\,\mathrm{Hz},\;\;T=450\,\mathrm{pK},
\end{equation}
a typical soliton size $\ell/2 = 0.9$~$\mu$m and 100~$^7$Li atoms, implying an oscillator frequency $\lambda_{\mathrm{osc}} = 1.2$~$\mu$m. For these parameters we find sufficiently short time scales of decoherence. While the other restrictions are fulfilled, the chosen parameters yield occupation probabilities $p_0 \approx 0.656$, $p_1\approx0.226$ and $p_2\approx0.078$ for the ground, first and second excited oscillator state. Hence, excited states of the CoM are expected to be significantly occupied. In Sec.~\ref{sec:excitations} we demonstrate that the signatures of quantum superposition states shown in Sec.~\ref{eq:highfidelity} are sufficiently robust against such excitations of the CoM wave function to higher lying oscillator states. An optimization of experimental parameters could be a topic for future research.

\section{Interference patterns for high-fidelity NOON-states}\label{eq:highfidelity}
In the following we exemplary investigate scattering bright solitons off very narrow delta-like and broader barrier potentials. At this point we neglect excitations of the CoM wave function. Within the effective potential approach we numerically investigate the CoM motion as described in Secs.~\ref{sec:effpot} and~\ref{sec:numimp}. The contrast of an interference pattern is defined as
\begin{equation}\label{eq:contrast}
C = \frac{I_{\mathrm{max}}-I_{\mathrm{min}}}{I_{\mathrm{max}}+I_{\mathrm{min}}},
\end{equation}
where $I_{\mathrm{max}}$ ($I_{\mathrm{min}}$) denotes the maximum (minimum) value of the intensity on a suitably chosen interval.

Time scales are given in units of the dimensionless time~$t/T_{\mathrm{osc}}$, where
\begin{equation}
T_{\mathrm{osc}} = 2\pi/\omega,
\end{equation}
and lengths scales in units of the CoM oscillator length
\begin{equation}\label{eq:losc}
\lambda_{\mathrm{osc}} = \left(\frac{\hbar}{Nm\omega}\right)^{1/2}.
\end{equation}

\begin{figure}[htb]
\begin{center}
\includegraphics[width = 1.0\linewidth]{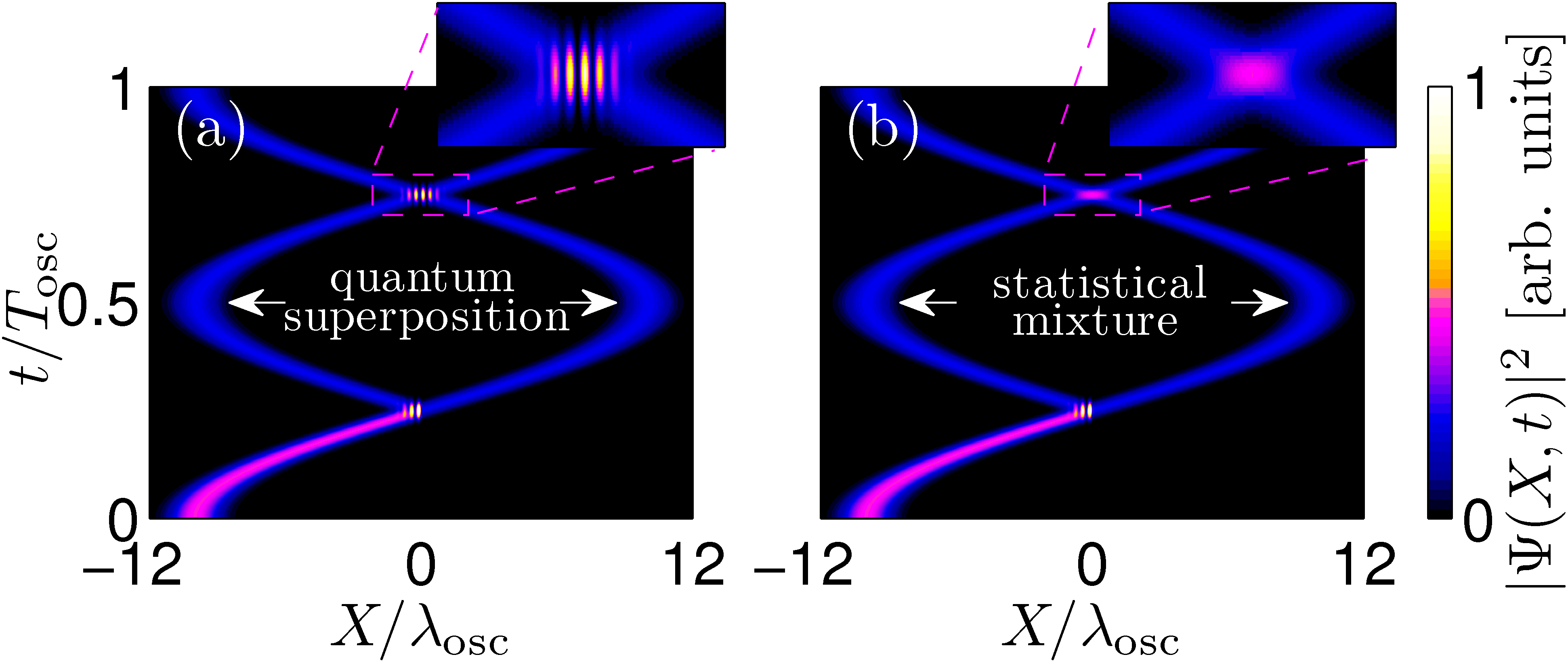}
\end{center}
\caption{\label{fig:qmstatdelta} (Color online) CoM density distribution $|\Psi\left(X,t\right)|^2$ for scattering a bright soliton off a delta-like effective potential~(\ref{eq:Veffdelta}), calculated on 6401 lattice points within the effective potential approach, cf.~Sec.~\ref{sec:effpot}. The potential height $v_0M\lambda_{\rm{osc}}/\hbar^2 = 1601$ is chosen such that 50\%-50\%-splitting takes place. The initial CoM density distribution corresponds to the shifted oscillator ground state centered around $X_0/\lambda_{\mathrm{osc}} = -10$. Variations of the initial particle number are modeled by a truncated Gaussian probability distribution~\cite{Grosspriv11} with mean value~$N=100$ and standard deviation~$\sigma = 5$ under consideration of particle numbers~$N\in[90,110]$. (a)~The scattering potential is switched off at $t/T_{\mathrm{osc}} = 0.5$, resulting in an interference of both parts of the superposition at about $t/T_{\mathrm{osc}} \approx 0.75$. The contrast~(\ref{eq:contrast}) of the interference pattern with fringe spacing~$0.37\lambda_{\rm{osc}}$ is $C\approx 1$ at $t/T_{\mathrm{osc}} \approx 0.75$. It is calculated on the interval $X\in\left[-0.5\lambda_{\rm{osc}}:0.5\lambda_{\rm{osc}}\right]$. (b)~As (a) but assuming that the quantum superposition is turned into a statistical mixture at~$t/T_{\rm{osc}}=0.5$.}
\end{figure}

First, we assume a delta-like effective potential~(\ref{eq:Veffdelta}). The time-evolution of the CoM density~(\ref{eq:CoMdensity}) is depicted in Fig.~\ref{fig:qmstatdelta}. Initially the CoM of the soliton is prepared in the shifted oscillator ground state. Scattering at the barrier potential in the middle of the harmonic confinement at approximately $t/T_{\rm{osc}}\approx 0.25$ leads to the creation of a nonlocal mesoscopic quantum superposition~(\ref{eq:NOON}). The barrier height throughout this work is chosen to ensure 50\%-50\%-splitting of the wave function. At $t/T_{\rm{osc}}=0.5$ the barrier potential is switched off, resulting in an interference of both parts of the wave function at $t/T_{\rm{osc}}\approx0.75$. The interference pattern with fringe spacing $0.37\lambda_{\rm{osc}}$ and maximal contrast $C\approx 1$ is stable under variations of the initial particle number, cf.~Fig.~\ref{fig:qmstatdelta}~(a). Assuming the quantum superposition to be turned into a statistical mixture results in a clearly different time-evolution as shown in Fig.~\ref{fig:qmstatdelta}~(b).

\begin{figure}[htb]
\begin{center}
\includegraphics[width = 1.0\linewidth]{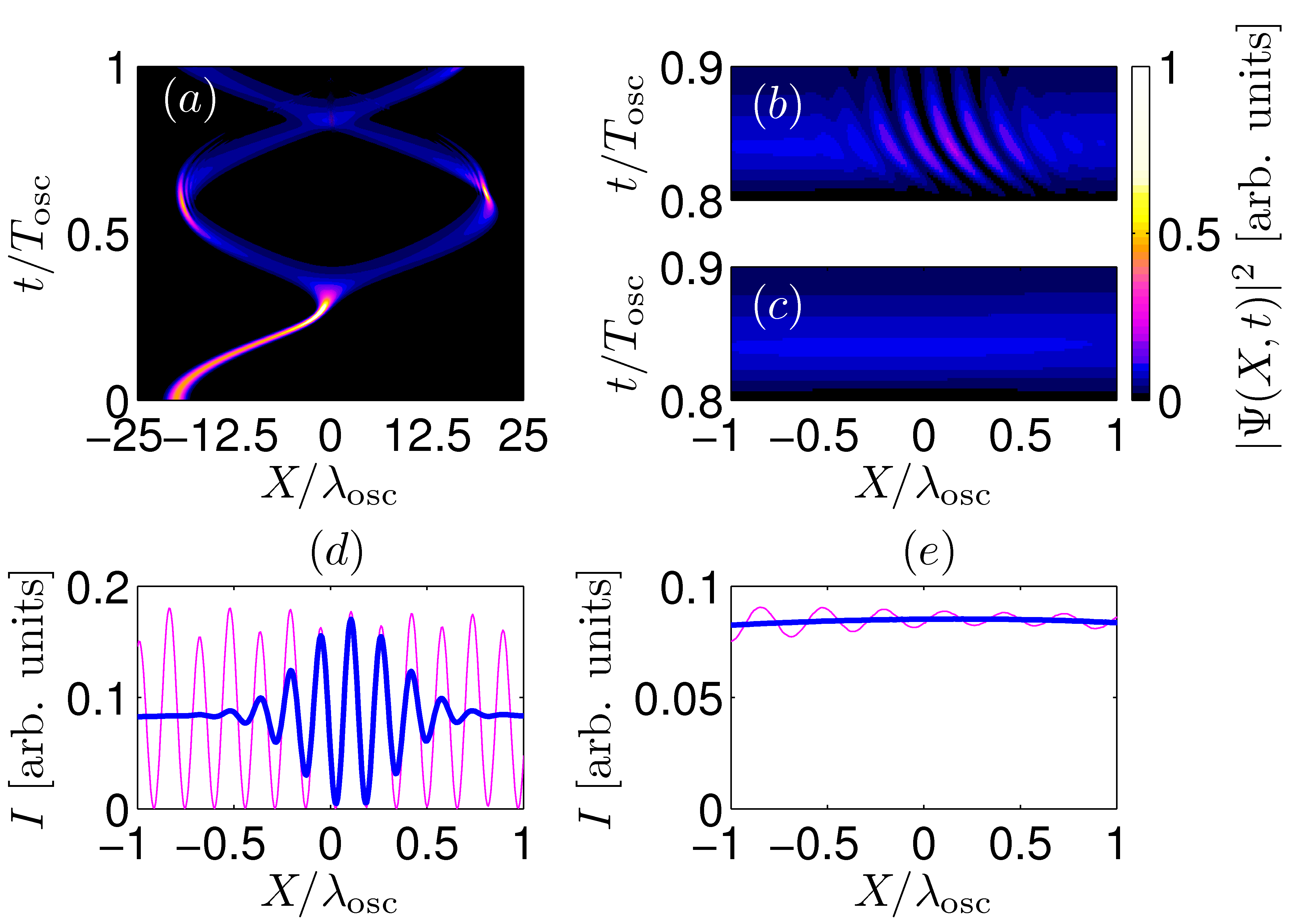}
\end{center}
\caption{\label{fig:filt} (Color online) CoM density distribution $|\Psi\left(X,t\right)|^2$ for scattering a bright soliton off a barrier potential~(\ref{eq:effpotU0}) of width $\ell/\lambda_{\rm{osc}} = 4$, calculated on 9601 lattice points within the effective potential approach from Sec.~\ref{sec:effpot}. The scaled potential height $U_0M\lambda^2_{\rm{osc}}/\hbar^2 = 200.4$ is chosen to ensure 50\%-50\%-splitting. The initial CoM density distribution corresponds to the shifted oscillator ground state centered around $X_0/\lambda_{\mathrm{osc}} = -20$. Initial variations of the particle number are modeled as in Fig.~\ref{fig:qmstatdelta}. (a)~The scattering potential is switched off at $t/T_{\mathrm{osc}} = 0.5$, resulting in an interference of both parts of the superposition around $t/T_{\mathrm{osc}} \approx 0.85$. (b) Zoom into~(a) to highlight the resulting interference pattern. (c)~As (b) but assuming that the quantum superposition is turned into a statistical mixture at $t/T_{\mathrm{osc}} \approx 0.5$. (d)~Density distribution for the quantum superposition at $t/T_{\mathrm{osc}}=0.85$. Blue, thick line: data as in~(a). Magenta, thin line: data for~$N=100$. The fringe spacing $0.16\lambda_{\rm{osc}}$ of the interference pattern is not affected by particle number variations. The maximal contrast $C\approx 1$ is reached for $N=100$ (calculated on the interval $\left[-0.5\lambda_{\rm{osc}}:0.5\lambda_{\rm{osc}}\right]$), while it is only slightly reduced to $C\approx0.95$ when particle number variations are included. (e)~Same as~(d) for the statistical mixture.}
\end{figure}

The same is depicted in Fig.~\ref{fig:filt} for a broader effective potential~(\ref{eq:Veffdelta}) with $\ell/\lambda_{\rm{osc}} = 4$ and a larger shift of the initial wave function. Also in this case the interference pattern with high contrast~$C\approx 0.95$ (again under consideration of variations of the initial particle number) can be used to clearly distinguish the quantum superposition state from a statistical mixture. The larger CoM kinetic energy results in a smaller fringe spacing in the interference pattern. The fringe spacing can in general be enhanced by choosing a smaller trapping frequency or smaller particle numbers. 
\subsection{Plane wave approximation of interference patterns}
\begin{figure}[htb]
\begin{center}
\includegraphics[width = 1.0\linewidth]{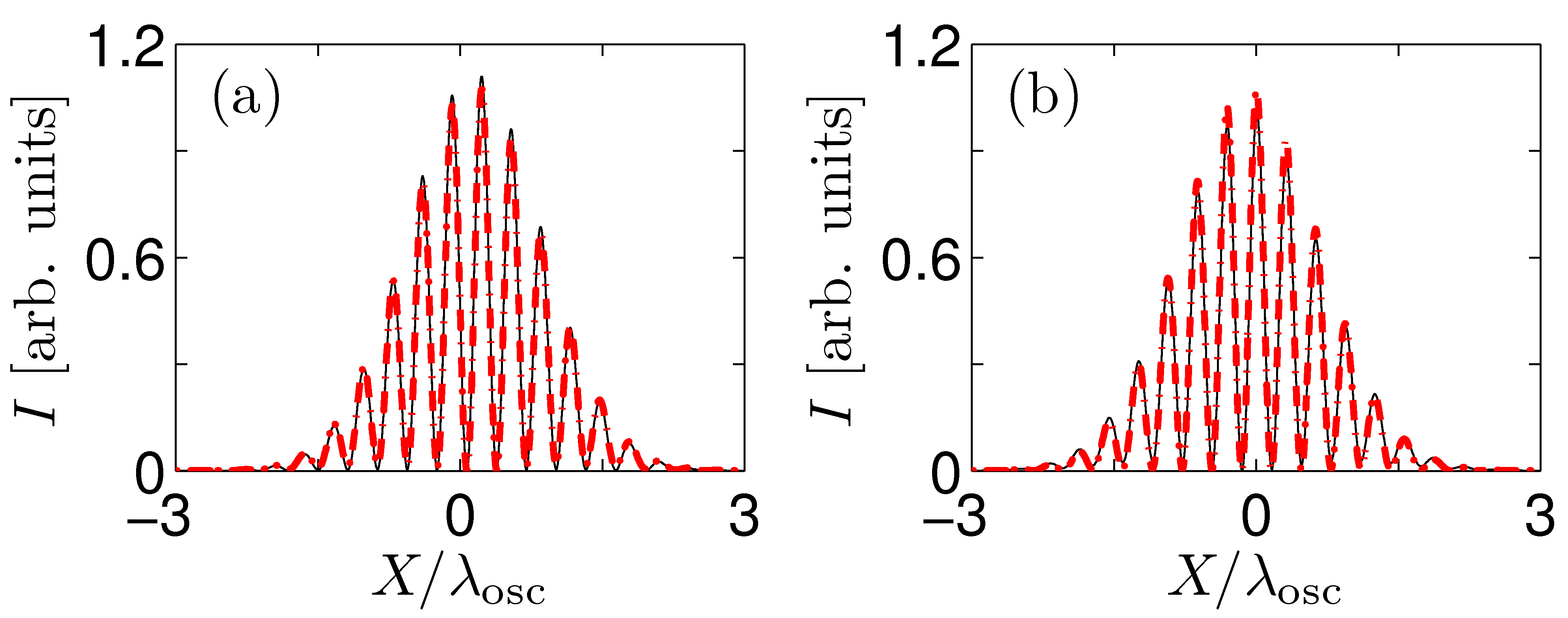}
\end{center}
\caption{\label{fig:deltaint} (Color online) Interference patterns after scattering a bright soliton off a barrier potential for 50\%-50\%-splitting of the wave function: numerical data (solid black line) and plane wave approximation (dashed red line). The approximate interference patterns are calculated according to~(\ref{eq:intpatwa}) with reflection and transmission coefficients for the potentials~(\ref{eq:Veffdelta}) and~(\ref{eq:Veffcoshshift}), respectively. The initial CoM wave function is the shifted oscillator ground state centered around $X_0/\lambda_{\mathrm{osc}}=-10$, resulting in the mean momentum~$\Omega\lambda_{\mathrm{osc}}=10$. (a)~Scattering at a shifted delta potential~(\ref{eq:Veffdelta}) with $X_{\mathrm{s}}/\lambda_{\mathrm{osc}}=-0.15$. (b)~Scattering at the shifted potential~(\ref{eq:Veffcoshshift}) with $\ell/\lambda_{\mathrm{osc}}=0.125$ and $X_{\mathrm{s}}/\lambda_{\mathrm{osc}}=-0.075$.}
\end{figure}

The numerically observed interference patterns in the CoM coordinate are well described in an approximation with plane waves, cf.~Fig.~\ref{fig:deltaint}.

For the scattering of a plane wave with wave-vector~$K$ at the delta-potential~(\ref{eq:Veffdelta}) the resulting wave function reads~\cite{Fluegge90}
\begin{eqnarray} \label{eq:wavefunc2}
\Psi_K(X)=\begin{cases}
  \mathrm{e}^{\mathrm{i}KX} + r_{\delta}\mathrm{e}^{-\mathrm{i}K X}  & \text{for }X<X_{\mathrm{s}}\\
  t_{\delta}\mathrm{e}^{\mathrm{i}KX} & \text{for }X>X_{\mathrm{s}},
\end{cases}
\end{eqnarray}
with reflection and transmission coefficients
\begin{equation}
r_{\delta} = \frac{\Omega}{\mathrm{i}K-\Omega},\;\;\; t_{\delta}= \frac{\mathrm{i}K}{\mathrm{i}K-\Omega}.
\end{equation}
In order to realize 50\%-50\%-splitting the condition 
\begin{equation}
\Omega \equiv K
\end{equation}
must be fulfilled.

The leading order behavior is understood by considering only the mean momentum component. The envelope of the interference pattern is captured by considering a wave packet centered around the mean momentum component~$K=\Omega$. To mimick the time-evolution induced by the harmonic confinement we assume for each component
\begin{equation}
u_K(X)=r_{\delta}\mathrm{e}^{\mathrm{i}KX}+t_{\delta}\mathrm{e}^{-\mathrm{i}KX}\mathrm{e}^{-2\mathrm{i}KX_{\mathrm{s}}},
\end{equation}
where we include shifts of the barrier potential out of the middle of the harmonic confinement. For a wave packet with Gaussian envelope this results in
\begin{equation}\label{eq:intpatwa}
\Psi(X) \propto\int_{-\infty}^{\infty}\mathrm{d}K u_K(X)\mathrm{e}^{-\frac{\left(K-\Omega \right)^2}{2}\lambda^2_{\mathrm{osc}}}.
\end{equation}
The absolute square $\left|\Psi(X)\right|^2$ then gives a good approximation of the interference pattern, cf.~Fig.~\ref{fig:deltaint}~(a).

In Fig.~\ref{fig:deltaint}~(b) it can be seen that the approximation works also for scattering at a shifted potential~(\ref{eq:effpotU0}),
\begin{equation}\label{eq:Veffcoshshift}
\tilde{V}_{\mathrm{eff,c}}(X) = \frac{U_0}{\cosh^2\left(\left(X+X_{\mathrm{s}}\right)/\ell\right)},
\end{equation}
with reflection and transmission coefficients given in~\cite{Landau00vol3}. The approximation works well for not-too-large shifts~$X_{\mathrm{s}}$ and not-too-broad potentials.

\section{Excitations in center-of-mass coordinate}\label{sec:excitations}
For the creation of quantum superposition states the restriction $k_\mathrm{B}T\ll |\mu_0|$ has to be fulfilled~\cite{WeissCastin09}. What happens if the much more stringent temperature restriction $k_\mathrm{B}T\ll \hbar \omega$ is relaxed, such that the CoM of the gas may be in an excited state? 

Bose-Einstein condensates at finite temperature have been investigated in~\cite{SinatraEtAl01} with a classical field method. Here, we simulate effects of finite temperature by assuming a statistical occupation of higher lying oscillator states for the CoM wave function while we assume that the internal degrees of freedom are still described by the quantum soliton~(\ref{eq:groundstate}).

In the following a renormalized occupation probability for a finite number $M$ of included oscillator states is assumed:
\begin{equation}
p_n = \frac{\exp\left(-\frac{\hbar\omega}{k_{\mathrm{B}}T}n\right)}{\mathcal{Z}},\;\;\mathcal{Z} = \sum_{n=0}^{M}\exp\left(-\frac{\hbar\omega}{k_{\mathrm{B}}T}n\right).
\end{equation}

The statistically averaged CoM density distribution then is given by
\begin{equation}\label{eq:statavCoM}
I\left(X,t\right)_{\mathrm{CoM,\mathrm{st}}} = \sum_{j = 0}^{M}p_{j}I\left(X,t\right)_{\mathrm{CoM},j}.
\end{equation}
Here $p_j$ is the occupation probability~(\ref{eq:stateocc}) of the $j$-th occupied oscillator state and $I\left(X,t\right)_{\mathrm{CoM},j}$ the CoM density distribution of the $j$-th oscillator state.

\begin{figure}[htb]
\begin{center}
\includegraphics[width = 1.0\linewidth]{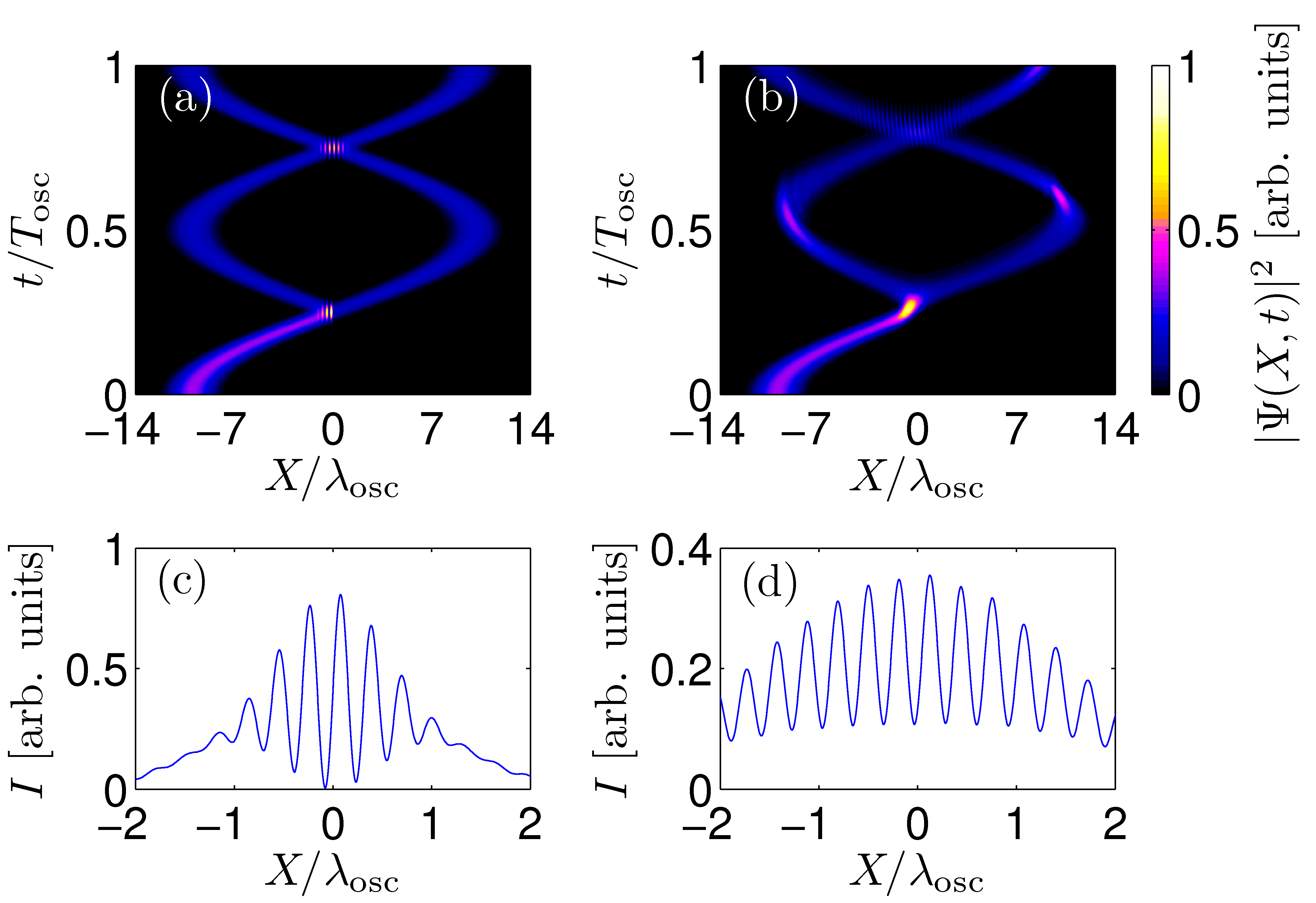}
\end{center}
\caption{\label{fig:excitationsdeltaint} (Color online) Same protocol as in Fig.~\ref{fig:qmstatdelta} \mbox{for~$N=100$}. Statistically averaged CoM probability density~(\ref{eq:statavCoM}) for \mbox{$T=450$~pK}, $\omega=2\pi\cdot10$~Hz, $N=100$~$^7$Li~atoms and~$M=2$. (a)~Delta-like barrier potential~(\ref{eq:Veffdelta}). The contrast at $t/T_{\mathrm{osc}}=0.75$ is reduced from $C\approx1$ to $C=0.987$. (b)~Scattering potential~(\ref{eq:effpotU0}) with $\ell = 1.5\lambda_{\mathrm{osc}}$. The contrast at $t/T_{\mathrm{osc}}=0.8$ is reduced from $C\approx1$ to~$C=0.574$. (c)~CoM-density from~(a) at $t/T_{\mathrm{osc}} = 0.75$. (d)~CoM-density from~(b) at~$t/T_{\mathrm{osc}} = 0.8$.}
\end{figure}

How are the interference patterns from section~\ref{eq:highfidelity} affected by such excitations? For a very narrow, delta-like effective potential~(\ref{eq:Veffdelta}) this is illustrated in figure~\ref{fig:excitationsdeltaint}~(a), where the averaged density distribution~(\ref{eq:statavCoM}) is displayed for $M=2$ and the parameters~(\ref{eq:expparam}). The contrast~(\ref{eq:contrast}) of the resulting interference pattern is only slightly reduced from one to $C=0.987$. For the experimentally realistic parameters~(\ref{eq:expparam}), corresponding to a width~$\ell = 1.5\lambda_{\rm{osc}}$ of the effective potential~(\ref{eq:effpotU0}), the results are displayed in Fig.~\ref{fig:excitationsdeltaint}~(b). The contrast~(\ref{eq:contrast}) is reduced from approximately one to $C=0.574$, a value still allowing a clear distinction from a statistical mixture. 

\begin{figure}[htb]
\begin{center}
\includegraphics[width = 1.0\linewidth]{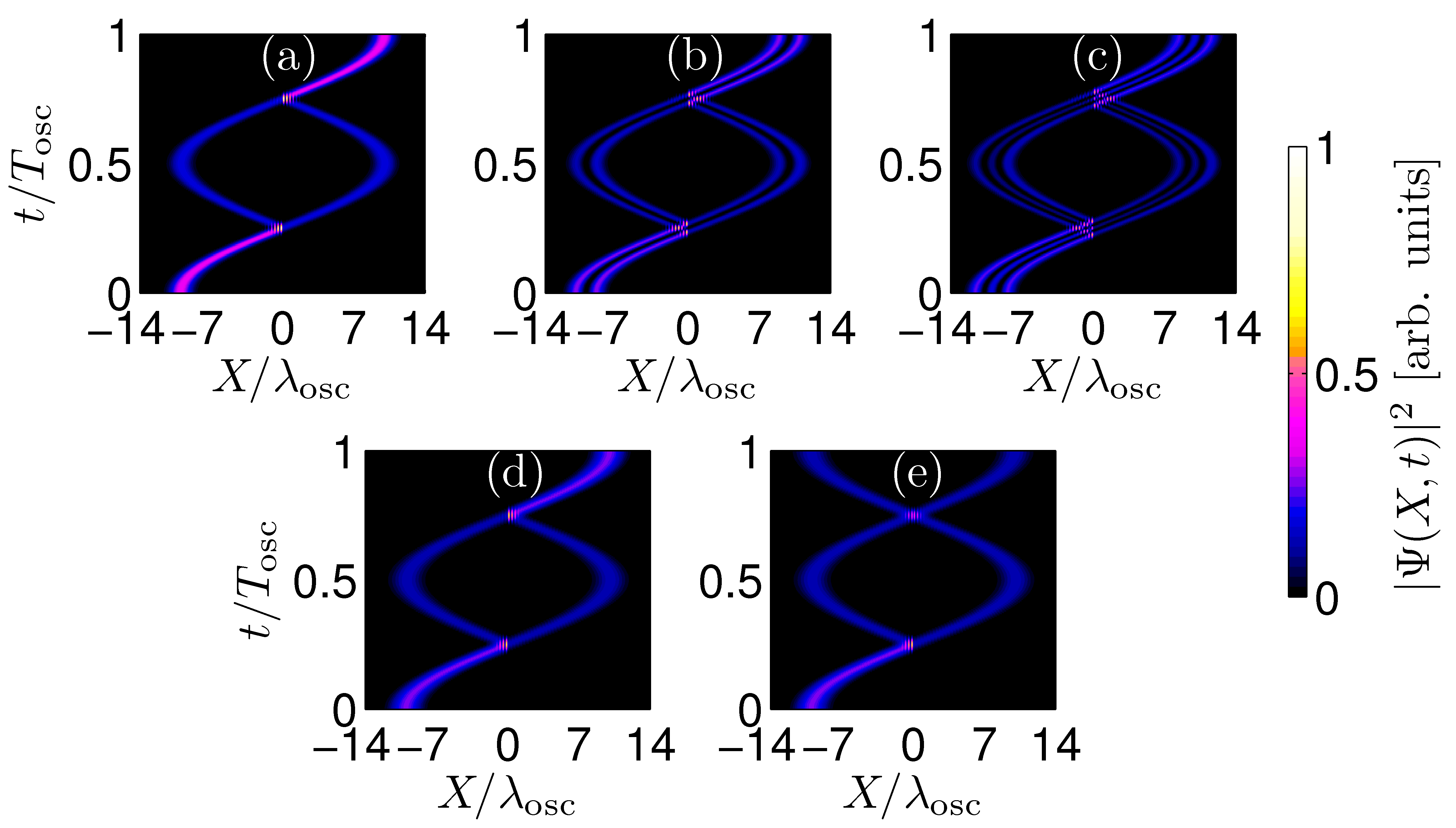}
\end{center}
\caption{\label{fig:excitationsdeltalr} (Color online) CoM probability density~$|\Psi\left(X,t\right)|^2$ for scattering twice off a delta potential~(\ref{eq:Veffdelta}). (a)~Initial state: ground state. The probability to find the particles to the right side of the barrier potential at $t/T_{\mathrm{osc}}=1$ is~$p_{\mathrm{right}}(T_{\mathrm{osc}})=0.985$. (b)~Initial state: first excited state. (c)~Initial state: second excited state. (d)~Statistically averaged CoM probability density~(\ref{eq:statavCoM}) for same parameters as in Fig.~\ref{fig:excitationsdeltaint}. The probability to find the particles to the right side of the barrier potential stays about the same: $p_{\mathrm{right}}(T_{\mathrm{osc}})=0.984$. (e)~Same as (d) but assuming that the quantum superposition is turned into a statistical mixture at $t/T_{\mathrm{osc}}\approx 0.5$.}
\end{figure}

In~\cite{GertjerenkenBillamEtAl12} another protocol, particularly suited for not-too-broad effective potentials, was investigated: Leaving the barrier potential switched on after the creation of the NOON-state~(\ref{eq:NOON}) and scattering twice off the barrier potential in a Mach-Zehnder like set-up leads to a particularly well-observable signature of quantum superposition states. We again assume a delta-like effective potential~(\ref{eq:Veffdelta}). For the CoM wave function initially in the ground state of the harmonic confinement the time-evolution is displayed in Fig.~\ref{fig:excitationsdeltalr}~(a): After one oscillation period all the particles would be found with a probability $p_{\mathrm{right}}(T_{\mathrm{osc}})\approx 1$ to the right of the barrier potential, if initially situated to the left side of the barrier potential\footnote{This behavior can also be understood in an approximation with plane waves~\cite{GertjerenkenBillamEtAl12}. The probability $p_{\mathrm{right}}(T_{\mathrm{osc}})$ is sensitive on the width of the barrier potential and shifts of the barrier potential out of the middle of the harmonic confinement. This could be used in an interferometric application to measure small potential gradients along the center of the harmonic trap, cf.~Figs.~2 and~3 from~\cite{GertjerenkenBillamEtAl12}.} (including variations of the initial particle number as in Fig.~\ref{fig:qmstatdelta} the value is $p_{\mathrm{right}}(T_{\mathrm{osc}}) = 0.985$). In the case of a statistical mixture the particles would be found with equal probability at either side of the barrier potential, allowing a clear distinction between quantum superposition and statistical mixture in a series of measurements. Figure~\ref{fig:excitationsdeltalr}~(b) and~(c) show the time-evolution for the first and second excited eigenstate. The statistically averaged density distribution~(\ref{eq:statavCoM}) is displayed in Fig.~\ref{fig:excitationsdeltalr}~(d) for $M=2$ included oscillator states and parameters~(\ref{eq:expparam}). This yields a statistically averaged probability $\langle p_{\mathrm{right}}(T_{\mathrm{osc}}) \rangle_{\mathrm{st}}\approx 0.984$ to find the particles to the right of the barrier potential. Figure~\ref{fig:excitationsdeltalr}~(e) shows the averaged CoM density distribution when the quantum superposition is turned into a statistical mixture at $t/T_{\mathrm{osc}}\approx0.5$: As at zero temperature the particles would be found with equal probability to either side of the barrier potential.

For a broader effective potential with $\ell = 1.5\lambda_{\mathrm{osc}}$ the probability to find the particles to the right of the barrier potential is $p_{\mathrm{right}}(T_{\mathrm{osc}})=0.230$ for the quantum superposition and $p_{\mathrm{right}}(T_{\mathrm{osc}})=0.115$ for the statistical mixture (cf.~Fig.~3 from~\cite{GertjerenkenBillamEtAl12}.). For the parameters~(\ref{eq:expparam}) we obtain $p_{\mathrm{right}}(T_{\mathrm{osc}})=0.169$ and $p_{\mathrm{right}}(T_{\mathrm{osc}})=0.085$. To allow for a good experimental distinguishability a large number of runs is required both for zero and finite temperature.

\section{Two-particle solitons: interference patterns for high- and low-fidelity quantum superposition states}\label{eq:generalsup}
Research on two-particle bound states includes experimental realizations~\cite{WinklerEtAl06} and theoretical investigations~\cite{PiilMolmer07,PetrosyanEtAl07,JavanainenEtAl10,Weiss10,SoerensenEtAl12,GertjerenkenEtAl11,Fogarty13}. The creation of the two-particle NOON-state
\begin{equation}\label{eq:cat2parti}
|\Psi_{\mathrm{NOON},2}\rangle = \frac{1}{\sqrt{2}}\left(|2,0\rangle + \mathrm{e}^{\mathrm{i}\phi}|0,2\rangle\right).
\end{equation}
via scattering at a barrier potential has been demonstrated numerically~\cite{Weiss10,GertjerenkenEtAl11,Fogarty13}. The subsequent recombination of both parts of the wave function again gives rise to an interference pattern~\cite{GertjerenkenEtAl11,Fogarty13} that is particularly well-observable in the CoM density, cf.~\cite{GertjerenkenEtAl11} and footnote~\ref{CoMandsingleparticledensity}.

The former considerations for high-fidelity NOON-states are extended in the following: we show that lowering the interparticle interaction can lead to the generation of low-fidelity quantum superposition states
\begin{equation} \label{eq:generalcats2}
|\Psi_{\mathrm{lowfid}}\rangle = \frac{1}{\sqrt{1+|a|^2+|b|^2}}\left(|2,0\rangle + a|1,1\rangle + b |0,2\rangle  \right)
\end{equation}
with complex coefficients $a$ and $b$. The occupation probabilities of the states $|n,N-n\rangle$, $n = 0,1,2$ are denoted $\tilde{p}_n$ in the following. We show that an interference pattern can still be visible in this generalized situation, even for strong deviation from the two-particle NOON-state~(\ref{eq:cat2parti}). A high ratio of kinetic energy to interaction energy finally leads to the creation of product states, cf. the observed behavior for $N=4$~\cite{GertjerenkenBillamEtAl12}: for non-interacting particles the set-up corresponds to the action of a single-particle beam-splitter, yielding occupation probabilities~$\tilde{p}_0=0.25$, $\tilde{p}_1=0.5$ and $\tilde{p}_2=0.25$.

Effects of the interparticle interaction have also been investigated in~\cite{Fogarty13} with a focus on single-particle densities.  Along the lines of~\cite{GertjerenkenEtAl11} we focus in the following in particular on interferences in the CoM density.

\subsection{Numerical results}
We investigate scattering a two-particle soliton off a delta-like barrier potential, situated in the middle of an additional harmonic confinement. Lengths scales are given in units of
\begin{equation}\label{eq:losc2}
\lambda_{\mathrm{osc,2}} = \left(\frac{\hbar}{2m\omega}\right)^{1/2}.
\end{equation}
The presented results are obtained via discretization of the two-particle Schr\"odinger equation with sufficiently small lattice spacing, cf.~Sec.~\ref{sec:numimp}. With the parameter $U/J$ the ratio of interparticle interaction strength to kinetic energy can be adjusted. A lattice with 201 sites has proven suitable to allow for a satisfying spatial resolution in reasonable computing time. The initial state, the two-particle ground state, is determined by imaginary time evolution~\cite{GlickCarr11}, on a lattice where the harmonic confinement is shifted 50 lattice sites to the left. For the real-time evolution the harmonic confinement is again centered around the middle of the lattice such that the initial state is situated on the edge of the harmonic confinement.
\begin{figure}[htb]
\begin{center}
\includegraphics[width = 1.0\linewidth]{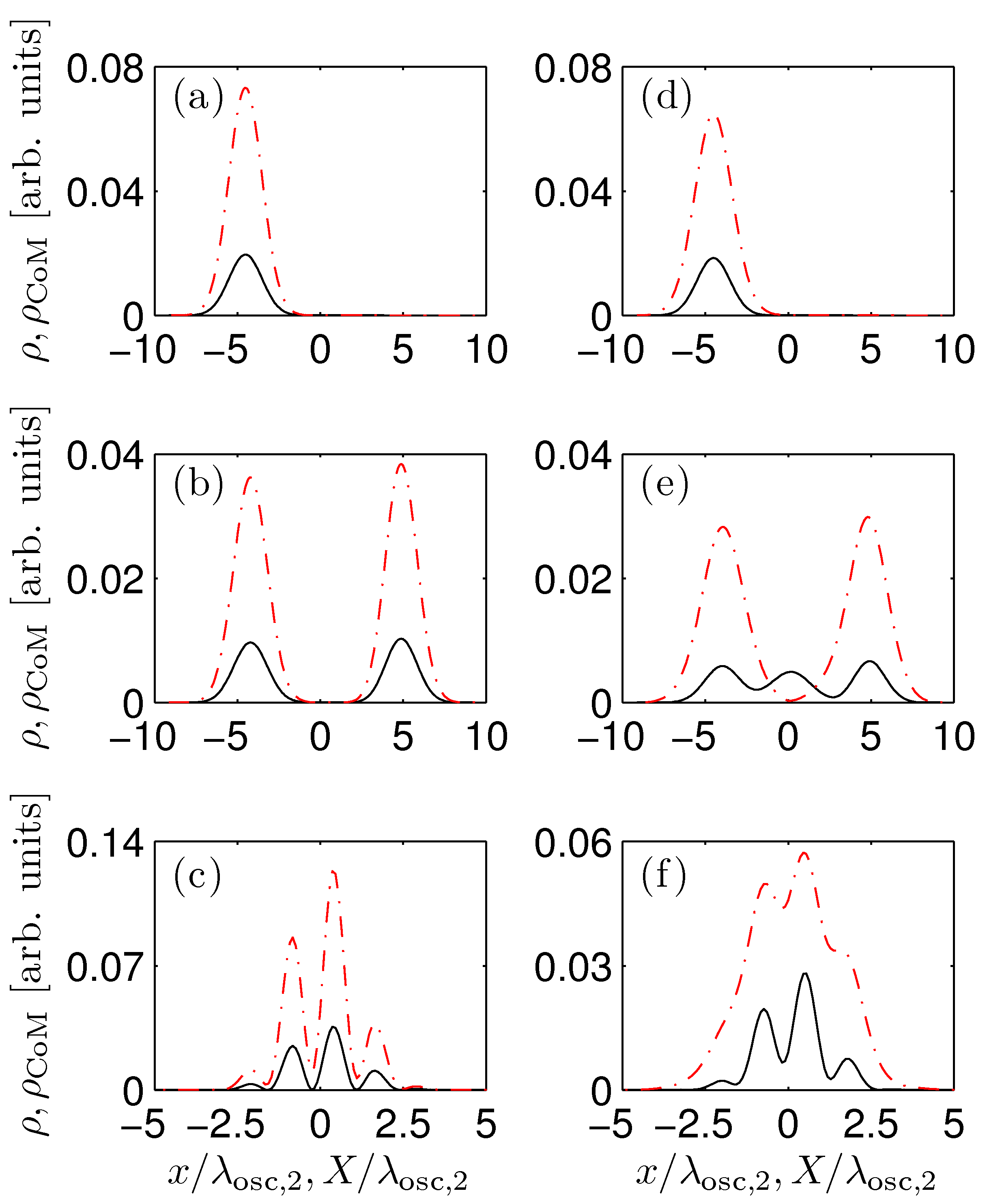}
\end{center}
\caption{\label{fig:kittenstimeevol} (Color online) Time evolution for a two-particle soliton being scattered off a delta potential in an additional slight harmonic confinement with \mbox{$A/J=9\cdot 10^{-6}$}, calculated on 201 lattice sites. The wave function is initially centered around $X/\lambda_{\mathrm{osc,2}} = - 4.61$. The initial state is determined by imaginary time evolution~\cite{GlickCarr11}. CoM density distribution $\rho_{\mathrm{CoM}}(X)$ (solid, black line) and single-particle density $\rho(x)$ (dash-dotted red line) versus spatial coordinate at exemplary times for interaction strengths $U/J=-1$ ((a)-(c)) and $U/J=-0.2$~\mbox{((d)-(f))}. (a)~Initial state. (b)~High-fidelity quantum superposition with occupation probabilities $\tilde{p}_0 \approx \tilde{p}_2 \approx 0.5$ and $\tilde{p}_1\approx 0$ at $t/T_{\mathrm{osc}} = 0.5$. (c)~Interference patterns at $t/T_{\mathrm{osc}}=0.77$. The contrast~(\ref{eq:contrast}) (calculated on the interval $-1.38<X/\lambda_{\mathrm{osc,2}}<1.38$) has a value of 1 for the CoM density distribution and is reduced to 0.71 in the single-particle density. \mbox{((d)-(f)):}~Same for $U/J=-0.2$. (e)~Resulting low-fidelity quantum superposition with $\tilde{p}_0 \approx 0.34$, $\tilde{p}_1 \approx 0.32$ and $\tilde{p}_2\approx 0.34$ at $t/T_{\mathrm{osc}} = 0.5$. (f)~Interference pattern at $t/T_{\mathrm{osc}} = 0.77$ with contrast~$C=0.83$ for the CoM density and $C=0.34$ for the single-particle density.}
\end{figure}

In Fig.~\ref{fig:kittenstimeevol} the time evolution of CoM density~(\ref{eq:CoMdensity}) and single-particle density~(\ref{eq:onedensity}) is compared for two different ratios of interaction strength to tunneling strength: for $U/J=-1$ panel~(a) shows the initial state localized to the left of the delta-like barrier potential. The scattering at the barrier potential at about $t/T_{\mathrm{osc}} \approx 0.25$ leads to the creation of a nonlocal two-particle NOON-state~(\ref{eq:cat2parti}). The strength of the scattering potential is chosen to ensure 50\%-50\%-splitting. The created NOON-state is depicted in Fig.~\ref{fig:kittenstimeevol}~(b) at $t/T_{\mathrm{osc}} = 0.5$, where two distinct peaks can be observed both for single-particle and CoM density. Removing the barrier potential gives rise to an interference of both parts of the wave function at $t/T_{\mathrm{osc}}\approx0.77$ as depicted in~Fig.~\ref{fig:kittenstimeevol}~(c). For the CoM density we obtain nearly perfect contrast~(\ref{eq:contrast}) while it is considerably reduced for the single-particle density~(cf.~\cite{GertjerenkenEtAl11}).

The same is depicted in Fig.~\ref{fig:kittenstimeevol}~(d)-(f) for the lower interaction strength~$U/J=-0.2$. While the single-particle density in~Fig.~\ref{fig:kittenstimeevol}~(e) still corresponds to a 50\%-50\%-probability to find the particles on either side of the potential, significant contributions of the state~$|1,1\rangle$ to the CoM density are observed, giving rise to the peak in the middle of the harmonic confinement. The contrast~(\ref{eq:contrast}) in the resulting single-particle density interference pattern again is considerably reduced, but an interference pattern in the CoM density is still clearly visible with contrast~$C=0.83$, cf.~Fig.~\ref{fig:kittenstimeevol}~(f).

\begin{figure}[htb]
\begin{center}
\includegraphics[width = 1.0\linewidth]{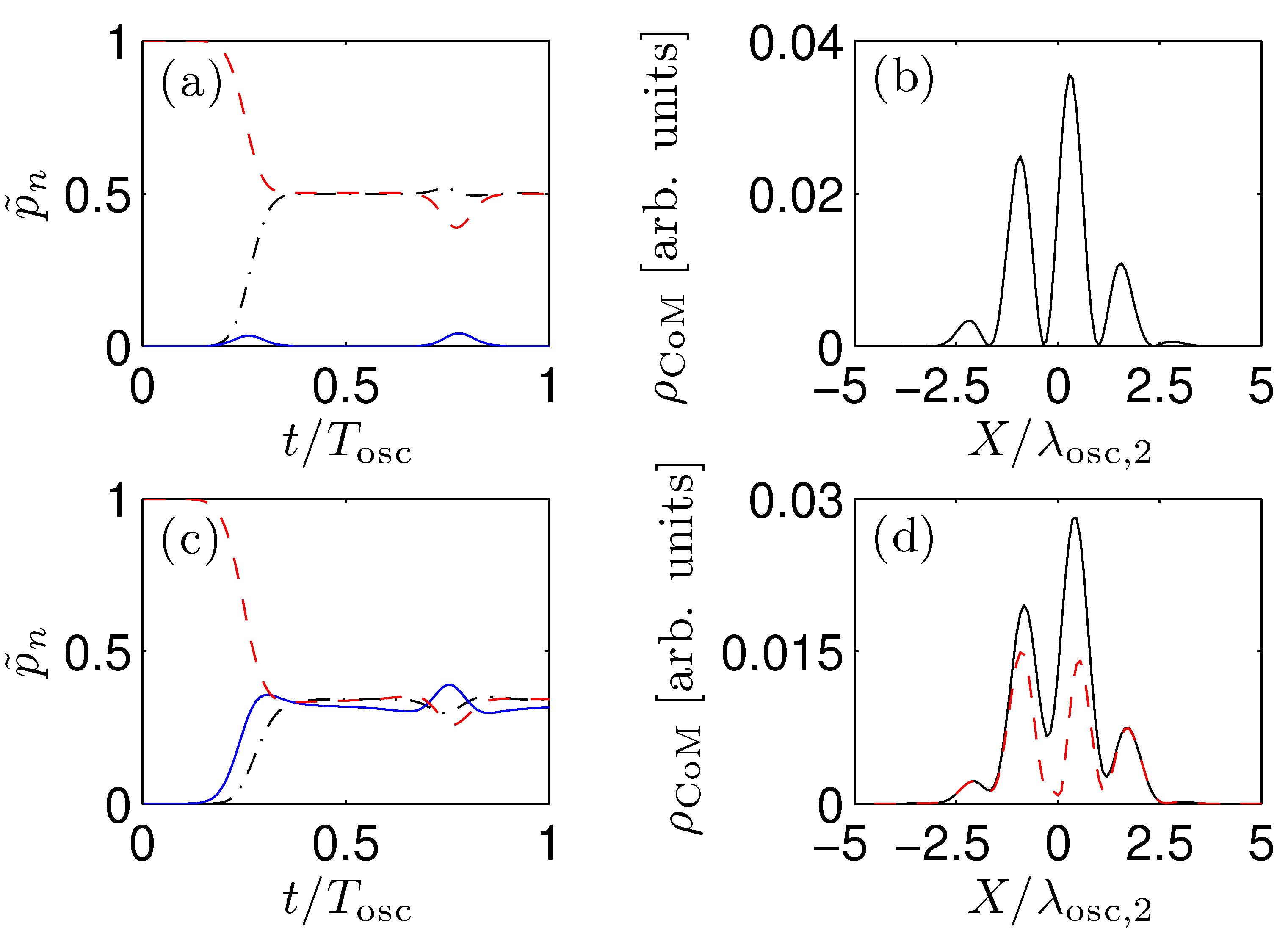}
\end{center}
\caption{\label{fig:kittensFockInt} (Color online) Same set-up as in Fig.~\ref{fig:kittenstimeevol}. Occupation probabilities $\tilde{p}_n$ as defined below Eq.~(\ref{eq:generalcats2}) versus dimensionless time $t/T_{\mathrm{osc}}$ for (a)~$U/J=-1$ and (c)~$U/J=-0.2$. Dash-dotted black line:~$\tilde{p}_2$, solid blue line:~$\tilde{p}_1$ and dashed red line:~$\tilde{p}_0$. (b)~Interference pattern in CoM density~$\rho_{\mathrm{CoM}}(X)$ for~$U/J=-1$ at $t/T_{\mathrm{osc}}\approx0.77$ with contrast~$C_{\mathrm{CoM}}=1.0$. No contribution from $|1,1\rangle$. (d)~Interference pattern in CoM density (solid black line) with~$C_{\mathrm{CoM}}=0.83$ and CoM density minus contribution from $|1,1\rangle$ (dashed red line) with~$C_{\mathrm{CoM,corr}}=0.90$ at $t/T_{\mathrm{osc}}=0.77$ for~$U/J=-0.2$.}
\end{figure}

For a further analysis, in the left panels of Fig.~\ref{fig:kittensFockInt} the time-evolution of the occupation probabilities~$\tilde{p}_n$ is displayed for interaction strengths $U/J=-1$ and $U/J=-0.2$. The worsening in contrast for lower interaction strength can be explained with the growing contribution~$\tilde{p}_1$. Removing it numerically can further improve the contrast of interference patterns in the CoM density, as displayed in Fig.~\ref{fig:kittensFockInt}~(d).

Experimentally, in principle, the suggested protocol could be investigated in lattices of double-wells, cf.~experiments like~\cite{CheinetEtAl08}: preparing the atoms initially in one of the wells and then switching off the short-wavelength laser, the two particles would be initially prepared on the edge of the remaining approximately harmonic confinement, leading to an oscillation in the potential well. A possibility to realize delta-like barrier potentials could be single atoms as scattering potentials.

\section{Conclusion}\label{eq:conclusion}
We have numerically investigated scattering bright solitons off barrier potentials with a focus on signatures of nonlocal high- and low-fidelity quantum superposition states. The chosen one-dimensional set-up with experimentally typical harmonic confinement gives rise to signatures that clearly distinguish quantum superposition states from statistical mixtures. The presented protocols naturally have the advantage that no opening of the harmonic trap is required, such that excitations due to trap opening~\cite{Castin09f} are avoided.

We use the mathematically rigorous~\cite{WeissCastin12} effective potential approach~\cite{WeissCastin09,SachaEtAl09} suitable for elastic scattering: switching off the barrier potential and recombining both parts of the NOON-state leads to high-contrast interference patterns in the CoM density both for narrow and broader barrier potentials. In combination with measurements confirming that all the particles are always clustered in a single lump with 50\% probability to either side of the barrier potential (to exclude single-particle effects), these interference patterns can serve as a clear indication of quantum superposition states. Another protocol particularly suited for narrow barriers is scattering twice off the barrier potential, cf. also~\cite{GertjerenkenBillamEtAl12}. While in general excitations can be a severe problem we have shown that both protocols are remarkably robust in this respect: finite temperature effects have been modeled by taking into account excitations of the CoM wave function to higher lying oscillator states.

For two-particle solitons and delta-like barrier potentials clear interference patterns -- again, particularly well-observable in the CoM density -- have been demonstrated not only for high- but also for low-fidelity quantum superposition states. While the presented results have been obtained for two particles a deduction of qualitatively similar results to higher particle numbers seems reasonable. This extends former considerations to interesting target states, advantageous in view of decoherence and allowing for shorter time-scales of the presented protocols by choosing higher initial CoM kinetic energies.

Despite in a different experimental regime than required for the proposed protocols, scattering bright quantum solitons off potential barriers is currently investigated experimentally~\cite{PollackEtAl10,Marchant2013}. The presented signatures could be used in future experiments to distinguish quantum superposition states from statistical mixtures. Due to the large number of data points required it seems advantageous to use arrays of one-dimensional tubes~\cite{GreinerEtAl01c}.

\section{Acknowledgements} 
I thank M.~Holthaus, M.~Oberthaler, A.~Streltsov, W.~Zurek and especially C.~Weiss for discussions. I acknowledge funding by the Studienstiftung des deutschen Volkes and the Heinz Neum\"uller Stiftung.


\end{document}